\documentclass[a4paper,11pt]{revtex4-1}
\usepackage{graphicx}
\usepackage{amsmath}
\usepackage{fancyhdr}
\usepackage{cancel}
\usepackage{color}
\setcitestyle{round}
\begin{document}
\title{A maximum entropy thermodynamics of small systems}
\author{Purushottam D. Dixit}
\affiliation{Biosciences Deptartment, Brookhaven National Laboratory}
\thanks{Corresponding author: Phone: (631) 344-3742; Email: pdixit@bnl.gov}
\begin{abstract}
We present a maximum entropy approach to analyze the state space of a small system in contact with a large bath e.g. a solvated macromolecular system. For the solute, the fluctuations around the mean values of observables are not negligible and the probability distribution $P(r)$ of the state space depends on the intricate details of the interaction of the solute with the solvent. Here, we employ a superstatistical approach: $P(r)$ is expressed as a marginal distribution summed over the variation in  $\beta$, the inverse temperature of the solute. The joint distribution $P(\beta, r)$ is estimated by maximizing its entropy. We also calculate the first order system-size corrections to the canonical ensemble description of the state space. We test the development on a simple harmonic oscillator  interacting with two baths with very different chemical identities viz. a) Lennard-Jones particles and b) water molecules. In both cases, our method captures the state space of the oscillator sufficiently well. Future directions and connections with traditional statistical mechanics are discussed.

\end{abstract}
\maketitle
\section{Introduction}
Recent developments in spectroscopic~\citep{roy2008practical} and single molecule manipulation techniques~\citep{greenleaf2007high,mehta1999single} allow direct measurements of very small systems. Examples of small systems from aqueous solution chemistry and biochemistry including proteins/nucleic acid molecules~\citep{garrettbiochemistry}, nanoparticles~\citep{rao2006chemistry}, and  hydrated ion [M(H$_2$O)$_{\rm n}$]$^{\rm k+}$ complexes interacting with bulk water solution~\citep{marcus1994simple}. Here, we are more interested in the description of the dynamics within the solute system, e.g. folding of the protein or interaction between nanoparticles, than the details of its interactions with the surrounding medium. In other words, we seek an {\it effective} thermodynamic description of the system  by integrating over the uninteresting `bulk' degrees of freedom. 

Due to the small size of the system and the high surface-to-volume ratio implies that the bath of  solvent particles cannot be treated as an ideal thermal bath, one that interacts only weakly with the system.  Consequently, the system-bath interactions have to be entertained at some detailed level of description. We expect that the behavior of a small system to be markedly different from systems with macroscopically large number of particles due to relatively heightened fluctuations in energy, volume, etc~\citep{hill,hill:nano}. 

To see this clearly, consider a large system coupled to a thermal bath. For simplicity, assume that the system can only exchange energy with the bath. Now imagine that the above system can be decomposed into subsystems {\bf A} and {\bf B} such that {\bf A} is very large compared to  {\bf B}. For convenience, we will identify {\bf A} as the solvent and {\bf B} as a solute e.g. a protein. Throughout this article, we will  use the terms `solute' and `system' and the terms `solvent' and `bath' interchangably. The energy $E$ of the composite system can be written as 
\begin{eqnarray}
E(r_{\bf A},r_{\bf B}) &\equiv& E_{\bf A}(r_{\bf A}) + E_{\bf B}(r_{\bf B}) + E_{\bf AB}(r_{\bf A},r_{\bf B}) \label{eq:decompEnergy}
\end{eqnarray}
where $E_{\bf A}$ ($E_{\bf B}$) is the interaction energy within the solvent (solute) and $E_{\bf AB}$ is the interaction between the solvent and the solute. $r_{\bf A}$ ($r_{\bf B}$) denote the collective coordinates of solvent (solute) particles. Note that $E \approx E_{\bf A} \gg E_{\bf B} (\sim E_{\bf AB})$. 

In the canonical ensemble description of the system at inverse temperature $\beta = 1/k_BT$, we write the distribution of states $P(r_{\bf A},r_{\bf B})$ as
\begin{eqnarray}
P(r_{\bf A},r_{\bf B}|\beta) &\propto& exp \left ( -\beta E(r_{\bf A},r_{\bf B}) \right ). \label{eq:canonical}
\end{eqnarray}

The marginal distribution for the solute degrees of freedom  is formally written as
\begin{eqnarray}
P(r_{\bf B}|\beta)\ &=& \sum_{r_{\bf A}} P(r_{\bf A},r_{\bf B}|\beta) \propto exp \left ( -\beta [E_{\bf B} (r_{\bf B}) +\phi_{\bf B}(\beta,r_{\bf B})] \right ). \label{eq:field}
\end{eqnarray}

The marginal distribution depends only on $r_{\bf B}$, the internal coordinates of the solute, and the inverse temperature $\beta$. Here, $\phi_{\bf B}(\beta,r_{\bf B})$ represents the {\it temperature dependent} effect of the solvent-solute interactions on the state space of the solute. Note that $\phi_{\bf B}(\beta,r_{\bf B})$ is a constant if the solute is very large, i.e. $E_{\bf B} \gg E_{\bf AB}$, and behaves independently of the solvent.  In this case, the solute acts as a thermodynamic system in its own right, in contact with an ideal thermal bath at the same inverse temperature (a thermal bath comprising of ideal gas particles).

Understanding the structure of the `molecular field' $\phi_{\bf B}(\beta,r_{\bf B})$ is of tremendous importance since $E_{\bf B}(r_{\bf B}) + \phi_{\bf B}(\beta,r_{\bf B})$ completely describes the phase space distribution of the solute~\citep{dixit:bj11,dixit2011elastic}. Even though Eq.~\ref{eq:field} is formally true, it has little practical value since $\phi_{\bf B}(\beta,r_{\bf B})$ depends in a non-trivial fashion on the details of the solute-solvent interaction $E_{\bf AB}$ and, in general, is quite hard to estimate~\citep{dixit2011thermodynamics,dixit:bj11}. {\it Ad hoc} assumptions about the structure of $\phi_{\bf B}(\beta,r_{\bf B})$ are commonplace in aqueous chemistry literature including (but certainly not limited to) examples such as the Generalized Born dielectric models~\citep{bashford2000generalized}, the non-linear Poisson-Boltzmann model~\citep{sharp1990calculating}.

Here, we present a maximum entropy approach to model the state space distribution $P(r_{\bf B})$ (see Eq.~\ref{eq:field}) by circumventing the problem of estimating the temperature dependent field $\phi_{\bf B}(\beta,r_{\bf B})$. We illustrate our framework, using molecular dynamics simulations, to study the state space of a Harmonic oscillator ({\bf B}) interacting with a) a bath of Lennard-Jones particles ({\bf A}) and b) a bath of water molecules ({\bf A}), two solvent media of completely different chemical identities. 

In sec.~\ref{sc:theory} we develop our method, in sec.~\ref{sc:limit} we explore connections with traditional statistical mechanics, and in sec.~\ref{sc:harmonic} we illustrate it for a harmonic oscillator coupled to two realistic baths. Finally in sec.~\ref{sc:disc}, we discuss future directions and possible connections with traditional statistical mechanics.

\section{Theory\label{sc:theory}}

\subsection{Maximum entropy superstatistics}

Consider a macromolecular solute interacting with a solvent as described above. The maximum entropy  (maxEnt) interpretation views the problem of determining the distribution $P(r_{\bf A},r_{\bf B})$ of the coordinates $\{ r_{\bf A}, r_{\bf B} \}$ of the composite system as an inference problem: $P(r_{\bf A}, r_{\bf B})$ is estimated from the {\it limited available knowledge} of the system~\citep{it_theory1}.  Briefly, the probabilities $P(r_{\bf A}, r_{\bf B})$ is estimated by maximizing the entropy $S[P(r_{\bf A}, r_{\bf B})]$ subject to constraints that the distribution reproduces some experimentally observed quantities such as the mean energy $\bar E$. The constrained objective function is given by Eq.~\ref{eq:maxEnt}~\citep{it_theory1,Press2012}.
\begin{eqnarray}
S [P(r_{\bf A}, r_{\bf B})]&-&   \beta \left (\sum_{r_{\bf A}, r_{\bf B}} P(r_{\bf A}, r_{\bf B})E(r_{\bf A}, r_{\bf B})  - \bar E \right ) + \gamma \left( \sum_{r_{\bf A}, r_{\bf B}} P(r_{\bf A}, r_{\bf B})- 1\right ). \label{eq:maxEnt}
\end{eqnarray} Here, $\beta$ and $\gamma$ are the Lagrange multipliers and \begin{eqnarray}
S [P(r_{\bf A}, r_{\bf B})] &=& -\sum_{r_{\bf A},r_{\bf B}} P(r_{\bf A}, r_{\bf B}) \log P(r_{\bf A}, r_{\bf B}).
\end{eqnarray} 

The estimated distribution $P(r_{\bf A}, r_{\bf B}|\beta)$ has the maximum entropy amongst all candidate distributions $P^*(r_{\bf A},r_{\bf B})$ that reproduce the experimental constraints (here, average energy). $P(r_{\bf A}, r_{\bf B}|\beta)$ is parameterized by a {\it unique} inverse temperature $\beta$ and is given by Eq.~\ref{eq:canonical}.  The maximum entropy estimate depends solely on the average energy $\bar E$ and results in a correct predictive theory only for macroscopically large systems~\citep{it_theory1,Press2012}. This is due to the fact that in a macroscopically large system, the fluctuation in energy, $\langle E ^2 \rangle$ - $\langle E \rangle^2$,  is negligible compared to the mean $\langle E \rangle$~\citep{it_theory1}. Consequently, the higher moments of the energy distribution $\langle E^n \rangle$ ($n >1$) can be estimated from the knowledge of the mean.  A naive application of the maximum entropy principle for a small system such as the solute is bound to result in predictions that do not match with experiments. 
%

Note if the solute ${\bf B}$ is sufficiently large, the internal interactions within the solute will vastly outweigh the solute-solvent interactions. Here, the solvent can be  integrated with the thermal bath and its effect absent from the description of the solute except for setting the inverse temperature $\beta$. In this case, the entropy of the solute $S[P(r_{\bf B})]$ itself is maximized subject to constraining the average energy of the solute.  Thus, for sufficiently large ${\bf B}$, the effect of {\bf A} on the phase space of {\bf B} can be represented by a single number: with respect to any prediction about the solute {\bf B}, the equivalence $P(r_{\bf A},r_{\bf B}) \leftrightarrow P(r_{\bf B})$ involves minimal loss of information. For a  solute of intermediate size, we assume that the solvent can be equivalently represented by allowing the inverse temperature of the solute to fluctuate i.e. we work with the {\it ansatz}  $P(r_{\bf A},r_{\bf B}) \leftrightarrow P(\beta,r_{\bf B})$ rather than $P(r_{\bf A},r_{\bf B}) \leftrightarrow P(r_{\bf B})$. Instead of maximizing the entropy $S [P(r_{\bf A}, r_{\bf B})]$ of $P(r_{\bf A}, r_{\bf B})$~\citep{lee:pre2012}, we maximize the entropy of the joint distribution $P(\beta,  r_{\bf B})$. Thus, we maximize
\begin{eqnarray}
S[P(\beta, r_{\bf B})] &=& -\sum_{\beta, r_{\bf B}} P(\beta, r_{\bf B}) \log P(\beta, r_{\bf B}) = S[P(\beta)] + \sum_{\beta} P(\beta) S_{\bf B}(\beta) \label{eq:compEntropy}
\end{eqnarray}
where $S[P(\beta)]$ is the entropy of 
\begin{eqnarray}
P(\beta) &=&  \sum_{r_{\bf B}} P(\beta, r_{\bf B})
\end{eqnarray}
and 
\begin{eqnarray}
S_{\bf B}(\beta)&=& -\sum_{r_{\bf B}} P(r_{\bf B}|\beta)  \log P(r_{\bf B}|\beta) \end{eqnarray} is the entropy of the solute in contact with an ideal thermal bath at inverse temperature $\beta$ and
\begin{eqnarray}
 P(r_{\bf B}|\beta)&\propto& exp \left (-\beta E_{\bf B}(r_{\bf B})\right )\end{eqnarray} is the canonical ensemble probability distribution.  The ideal thermal bath can be constructed by replacing the solvent by a dilute ideal gas solvent {\bf A$'$} at the same temperature such that $E_{\bf B} \gg E_{\bf A'B}$.

In maximization, a straightforward choice for the constraint is $\bar E_{\bf B}$, the observed mean energy of the solute {\bf B}. Since the entropy of the joint distribution $P(\beta, r_{\bf B})$ is maximized rather than that of the marginal distribution $P(r_{\bf B})$, the measured entropy $\bar S_{\bf B}$ itself becomes a constraint~\citep{Crooks2008}. Thus, the constrained optimization function, including the Lagrange multipliers and summing over $r_{\bf B}$ degrees of freedom, is (from Eq.~\ref{eq:compEntropy}),

\begin{eqnarray}
S[P(\beta)] &+& \lambda \left ( \sum_{\beta} S_{\bf B}(\beta) P(\beta)- \bar S_{\bf B} \right )+  \gamma \left ( \sum_{\beta} P(\beta) - 1\right ) - \zeta \left ( \sum_{\beta} \langle E_{\bf B}\rangle_\beta P(\beta) - \bar E_{\bf B}\right ).
\end{eqnarray}
Here\begin{eqnarray}
\langle E_{\bf B} \rangle_\beta &=& \sum_{r_{\bf B}} E_{\bf B}(r_{\bf B}) P(r_{\bf B}|\beta)
\end{eqnarray} is the average energy of the solute when it is coupled to an ideal thermal bath ({\bf A}$'$) at an inverse temperature $\beta$. 

We have transformed the problem of maximizing the entropy of the joint distribution $P(\beta, r_{\bf B})$ to the one of maximizing the entropy of $P(\beta)$. Carrying out the maximization,
\begin{eqnarray}
P(\beta) &=& \frac{1}{Z(\lambda,\zeta)} exp \left ( \lambda S_{\bf B}(\beta) - \zeta \langle E_{\bf B} \rangle_{\beta} \right )  \label{eq:pbeta}
\end{eqnarray} where 
\begin{eqnarray}
Z(\lambda, \zeta) &=& \sum_\beta  exp \left ( \lambda S_{\bf B}(\beta) - \zeta \langle E_{\bf B} \rangle_{\beta} \right ) 
\end{eqnarray} is the generalized partition function. The marginal distribution $P(r_{\bf B})$ is written as
\begin{eqnarray}
P(r_{\bf B}) &=& \sum_{\beta} P(r_{\bf B}|\beta) P(\beta) \label{eq:pr}
\end{eqnarray}

Eq.~\ref{eq:pbeta} and Eq.~\ref{eq:pr} are the main theoretical results of this work.  In deriving Eq.~\ref{eq:pbeta} and Eq.~\ref{eq:pr}, we have assumed that the effect of the bulk solvent medium {\bf A} on the solute {\bf B} can be captured by allowing the inverse temperature of the solute to fluctuate. Note that the above coarse graining approach implies that the marginal distribution $P(r_{\bf B})$ belongs to a restricted family of distributions. We believe that this approach will be successful if the solute is of intermediate scale i.e. when $\langle \delta \phi^2 \rangle$ is small but cannot be completely neglected.

\subsection{Implicit solvation}

In the development presented here, notice that the predictions about the solute state space (Eq.~\ref{eq:pbeta} and Eq.~\ref{eq:pr}) are independent of the details of the solute-solvent interactions.   From the perspective of {\it effective} interaction models, the current development closely resembles an implicit solvation model.  

The framework assumes that solvent induced modulation in the state space of the solute is completely characterized by allowing the temperature of the solute to vary. The distribution $P(\beta)$ of the temperature $\beta$ of the solute is governed by $\lambda$ and $\zeta$. In short, $\lambda$ and $\zeta$ (along with $S_{\bf B}(\beta)$ and $\langle E_{\bf B} \rangle_\beta$; both are properties of the solute {\it only}) completely describes solute-solvent interactions. This is the result of the coarse graning where $P(r_{\bf A},r_{\bf B})$ is assumed to be equivalent to $P(\beta, r_{\bf B})$ with respect to all predictions about the solute {\bf B}. We believe that Eq.~\ref{eq:pbeta} and Eq.~\ref{eq:pr} present an implicit solvation model that is independent of the details of the solute-solvent chemstry and has a rigorous basis in the maximum entropy framework. We leave it for further studies to study more realistic systems using the current framework.

\section{Connections to statistical mechanics\label{sc:limit}}

It is instructive to examine the limiting behavior of  Eq.~\ref{eq:pbeta}. Let us first write $\zeta = \beta_0 \lambda$. The maximum of the $P(\beta)$ distribution can be found out by setting the first $\beta$ derivative of $\lambda S_{\bf B}(\beta) - \beta_0 \lambda \langle E_{\bf B} \rangle_{\beta}$ to zero. Differentiating with respect to $\beta$, we get
\begin{eqnarray}
-\lambda \frac{C_v(\beta)}{\beta} + \lambda \beta_0  \frac{C_v(\beta)}{\beta^2} &\equiv& 0 \nonumber 
\end{eqnarray}

It's easy to check that the second derivative is positive at $\beta=\beta_0$. Thus, if the heat capacity $C_v(\beta)$ and $\lambda$ are non-zero, the maximum of the $P(\beta)$ distribution lies at $\beta = \beta_0$. 

For large $\lambda$ keeping $\beta_0$ finite, $P(\beta)$ will tend to a narrowly peaked distribution around $\beta = \beta_0$ and can be described by a normal distribution. The width of the normal distribution is dictated by the second derivative. If $C_v(\beta)$ doesn't vary rapidly around $\beta = \beta_0$, we can show that for large $\lambda$, 
\begin{eqnarray}
P(\beta) &\sim& \mathcal N \left (\beta_0, \frac{\sqrt{2} \beta_0}{\sqrt{\lambda C_v(\beta_0)}} \right ).\label{eq:firstorder}
\end{eqnarray}

Thus, as $\lambda \rightarrow \infty$, the state space of the solute {\bf B} is described by a canonical ensemble distribution at inverse temperature $\beta = \beta_0 =\zeta/\lambda$. Note that as $\lambda \rightarrow \infty$, $\beta_0$ is also the temperature of the surrounding bath. We conclude that  $\lambda < \infty$ captures the strength of the solute-solvent interactions and $\beta_0 = \zeta/\lambda$ is the effective temperature of the solute.  

For weak  solute-solvent interactions, i.e. large $\lambda$, the first order corrections in solute behavior due to solute-solvent interactions, i.e. the first order estimate of $\phi(\beta_0,r_{\bf B})$ (see Eq.~\ref{eq:field}), can be computed if one knows the heat capacity of the solute at $\beta_0$, the effective temperature of the solute.   Dropping the $\beta_0$ dependence of $C_v$ for brevity and writing $P(r_{\bf B})$

\begin{eqnarray}
P(r_{\bf B}) &=& \int P(\beta) P(r_{\bf B}|\beta) d\beta \approx \int_0^\infty \mathcal N \left (\beta_0, \frac{\sqrt{2}\beta_0}{\sqrt{\lambda C_v(\beta_0)}} \right ) exp \left (-\beta E(r_{\bf B}) + \beta F(\beta_0) \right ) d\beta 
\end{eqnarray}

\begin{eqnarray}
	&\propto& exp \left ( -\beta_0 E(r_{\bf B}) \right )  exp \left ( \frac{ \beta_0^2 E(r_{\bf B})^2}{\lambda C_v} \right ) \left ( 1 + {\rm Erf} \left [ \frac{\lambda C_v - 2\beta_0E(r_{\bf B})}{2\sqrt{ C_v \lambda }}\right] \right ) \label{eq:correction}
\end{eqnarray}

\vspace{3mm}
From Eq.~\ref{eq:correction}, by expanding for energies small compared to $\lambda C_v$, we estimate $\phi(\beta_0,r_{\bf B})$  as (see Eq.~\ref{eq:field} for the definition of $\phi$)
\begin{eqnarray}
\phi(\beta_0,r_{\bf B}) &\approx& \frac{e^{-\frac{C_v \lambda }{4}} E(r_{\bf B}) }{ \sqrt{\pi C_v \lambda }}-\frac{\beta_0E(r_{\bf B})^2 }{C_v \lambda } \label{eq:phi}
\end{eqnarray}

As expected, $\phi(\beta_0,r_{\bf B}) \sim 0$ when $\lambda \rightarrow \infty$. It will be interesting to explore if Eq.~\ref{eq:correction} and Eq.~\ref{eq:phi} can be exploited in molecular dynamics simulations. We leave it for further studies.

Even under the simplifying condition of weak solute-solvent interactions, i.e. large $\lambda$, the maximum entropy approach suggests that the solvent modulates the phase space behavior of the solute in a highly non-trivial manner. Eq.~\ref{eq:correction} and Eq.~\ref{eq:phi} provide a glimpse into the mechanism of the solvent-induced modulation of the phase space of the solute.  The first term effectively {\it decreases} its temperature while the second term increases the propensity to sample high energy states thus effectively increasing the temperature.  

Notice that the phase space described by Eq.~\ref{eq:phi} is identical to a maxEnt probability distribution where the mean energy $\langle E \rangle$ {\it and} the fluctuation in the energy $\langle E^2 \rangle$ are constrained (see Eq.~\ref{eq:phi} and Eq.~\ref{eq:field}). Previously, the predictions from Eq.~\ref{eq:phi} have been validated for ferromagnetic materials~\citep{chamberlin2009beyond,chamberlin2009fluctuation}. In the current framework however, instaed of constraining the higher moments of energy, we have constrained the average entropy of the small system. The average entropy is a natural constraint since due to non-negligible system-bath interactions unlike for a thermodynamic system, the entropy of the distribution $P(r_{\bf B})$ of the phase space of the small system will not be at its maximum under the constraint of mean energy. We speculate that if the {\it ansatz} $P(r_{\bf A},r_{\bf B}) \leftrightarrow P(\beta,r_{\bf B})$ (see above) is violated, constraining the higher moments of the energy may prove useful.

\section{Harmonic oscillator\label{sc:harmonic}}

\subsection{Theory}

We analytically illustrate the above development (Eq.~\ref{eq:pbeta}, Eq.~\ref{eq:pr}, and Eq.~\ref{eq:phi}), using molecular dynamics simulations~\citep{namd:cc05},  for a harmonic oscillator, comprising of two Lennard-Jones particles, coupled to a solvent bath of a) hydrophobic Lennard-Jones particles and b) water molecules. 

Imagine a harmonic oscillator solute ({\bf B}) interacting  with a bath of solvent particles ({\bf A}) at inverse temperature $\beta$. Without loss of generality, let the energy of the oscillator be $E(r) = r^2$.  If the interactions of the oscillator with the surrounding particles are very weak, we know that the probability distribution $P(r)$ is parametrized by the inverse temperature and is given by
\begin{eqnarray}
P(r|\beta)&=&\frac{4 r^2 \beta ^{3/2} e^{-r^2 \beta }}{\sqrt{\pi }} \label{eq:pr0}.
\end{eqnarray}

Note that for the harmonic oscillator in contact with a weakly interacting bath,  the average energy $\langle E \rangle_{\beta} \sim 1/\beta$ and the entropy $S(\beta) \sim \log \beta$ upto an additive and multiplicative constant. If the oscillator-solvent interactions are not weak, Eq.~\ref{eq:pr0} does not adequately describe $P(r)$ (see Fig.~\ref{fg:expt}). In this case, as discussed above, we allow the inverse temperature $\beta$ of the oscillator to fluctuate. From Eq.~\ref{eq:pbeta},
\begin{eqnarray}
P(\beta) &\propto& exp \left ( \lambda \log \beta - \frac{\zeta}{\beta} \right ) \nonumber \\	
	&=& \frac{e^{-\frac{\zeta }{\beta }} \beta ^{-\lambda } \zeta ^{\lambda -1}}{\Gamma (\lambda -1)}. \label{eq:pbetaHO}
\end{eqnarray}

\begin{figure}
	\includegraphics[scale=0.9]{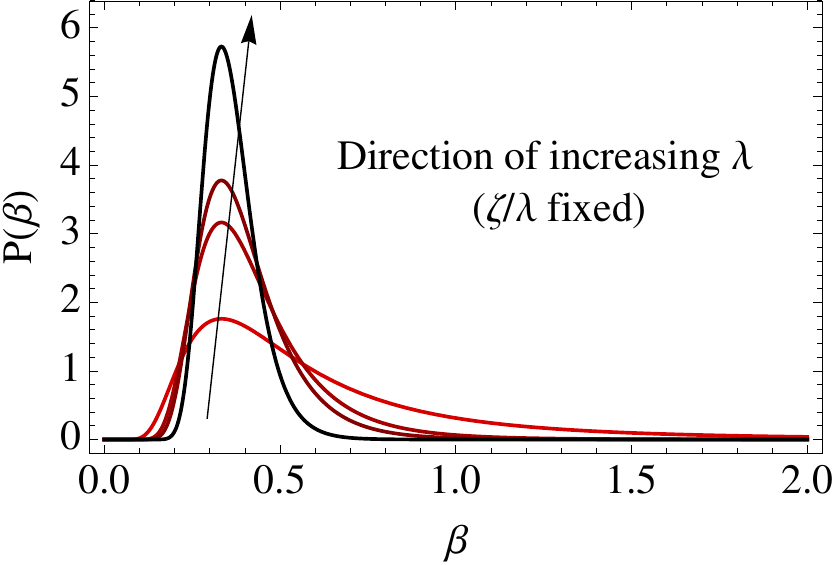}
	\caption{The distribution $P(\beta)$ of the temperature of the harmonic oscillator solute. As $\lambda \rightarrow \infty$, $P(\beta)$, the distribution of $\beta$ approaches a Dirac delta distribution. Notice that at small values of $\lambda$, the $P(\beta)$ distribution is very broad. Hence, while $\zeta/\lambda$ is equivalent to the temperature of the solute, $\lambda$ captures the strength of the interactions between the solvent and the solute.\label{fg:pbeta}}
\end{figure}

Before we derive the marginal distribution $P(r)$, let us inspect the limiting behavior of $P(\beta)$. Figure.~\ref{fg:pbeta} shows $P(\beta)$ for different values of $\lambda$ and $\zeta$, keeping $\beta_0= \zeta/\lambda$ fixed. Observe that at fixed $\beta_0$, $P(\beta)$ distributions with higher values of $\lambda$ is peaked and becomes a dirac delta distribution $\delta(\beta - \zeta/\lambda)$. Thus, as seen above, while $\beta_0 = \zeta/\lambda$ represents the temperature of the solute, $\lambda$ characterizes the strength of its interactions with the solvent: higher $\lambda$ implies weaker interactions. We hypothesize that in an experiment where the solvent composition is fixed, $\lambda$ cannot be changed smoothly while $\beta_0=\zeta/\lambda$ can be tuned by changing the temperature of the bath; however, in case of an aqueous solvent, $\lambda$ can be changed by perturbing the solute-solvent interactions using osmolytes and salts.

Finally, the marginal distribution $ P(r) = \int P(\beta) \cdot P(r|\beta) d\beta$ is given by,
\begin{eqnarray}P(r) &=& \frac{8 r^{\lambda -\frac{1}{2}} \zeta ^{\frac{\lambda }{2}+\frac{1}{4}} K_{\lambda -\frac{5}{2}}\left(2 r \sqrt{\zeta }\right)}{\sqrt{\pi } \Gamma (\lambda
   -1)} \label{eq:Pfinal1}
\end{eqnarray}
Here, $K_{\gamma}(x)$ is the modified Bessel function of the second kind with parameter $\gamma$.  To understand Eq.~\ref{eq:Pfinal1} physically, let's write $\zeta = \beta_0 \lambda$ and
calculate the average energy 
\begin{eqnarray}
\langle E \rangle &=& \langle r^2 \rangle = \frac{3 (\lambda -1)}{2 \beta_0 \lambda }\label{eq:moments}
\end{eqnarray}
As $\lambda \rightarrow \infty$,
\begin{eqnarray}
\langle E \rangle &=& \frac{3}{2\beta_0}\end{eqnarray}

Hence, as $\lambda \rightarrow \infty$, the oscillator behaves as if it is in contact with an ideal thermal bath at inverse temperature $2\beta_0/3$. Similarly, it is easy to check that the limiting behavior of Eq.~\ref{eq:Pfinal1} is equal to that of Eq.~\ref{eq:pbeta}, the distribution of states of a harmonic oscillator in contact with an ideal thermal bath.


\subsection{Numerical simulation}

An interesting aspect of the development above is that it hypothesizes that the form of the probability distribution $P(r_{\bf B})$ independent of the chemistry of the solute-solvent interactions. The solvent medium affects the solute only through the `coupling parameter' $\lambda$ (see Eq.~\ref{eq:pbeta} and Eq.~\ref{eq:pr}). We test this prediction by studying a harmonic oscillator interacting  with two systems that are chemically very different from each other viz. a bath of Lennard-Jones particles and a bath of water molecules (see appendix for details).

Fig.~\ref{fg:expt} shows the observed probability distribution (black circles) from the molecular dynamics simulations, the best-fit canonical ensemble distribution (black line, see Eq.~\ref{eq:pr0}), and the best-fit of Eq.~\ref{eq:Pfinal1} (see appendix for the fitting procedure).  Note that even though these two solvent systems have very different chemical identities, fig.~\ref{fg:expt} clearly shows that the empirically observed distribution for the two systems has an extended tail that cannot be captured by the canonical ensemble distribution. Meanwhile, the distribution $P(r)$ of Eq.~\ref{eq:Pfinal1} captures the empirically observed distribution extremely well.

\begin{figure*}
	\includegraphics[scale=0.95]{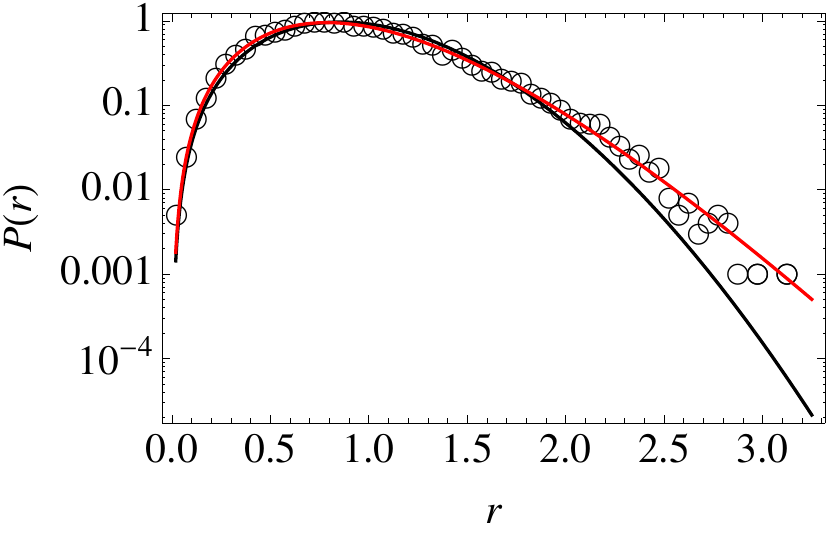}
	\includegraphics[scale=0.95]{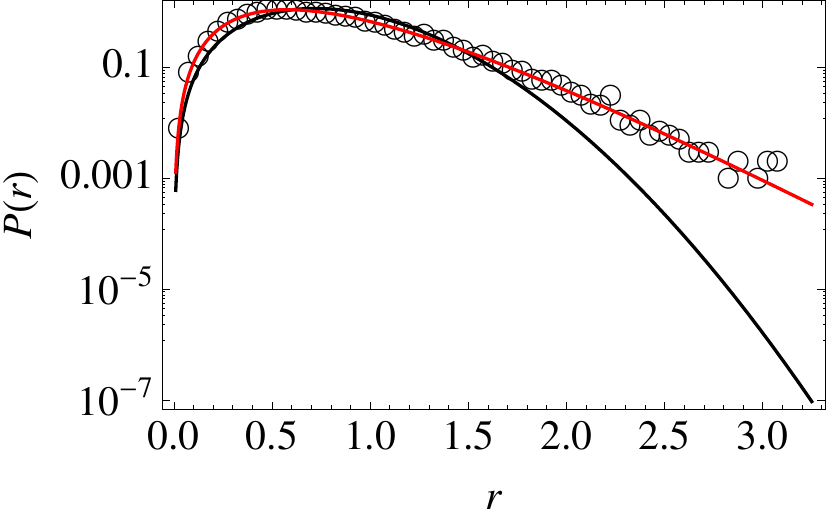}
	\caption{The experimentally observed $P(r)$ distribution (black circles) compared with the best-fit canonical ensemble distribution (black line, Eq.~\ref{eq:pr0}) and the best-fit of Eq.~\ref{eq:Pfinal1} (red line). Left panel shows the harmonic oscillator interacting with water molecules and the right panel shows the oscillator interacting with Lennard-Jones particles. Note that these two solvent systems have very different chemical identities. Yet, Eq.~\ref{eq:Pfinal1} describes the experimentally observed distribution over 5 orders of magnitude especially in the extended tail that the canonical ensemble distribution Eq.~\ref{eq:pr0} fails to capture. \label{fg:expt}}
\end{figure*}

\section{Discussion\label{sc:disc}}



In traditional statistical mechanics, the inverse temperature $\beta$ is an intensive variable and its fluctuation does not have a direct physical interpretation. Here, the entropy is related to heat loss and is a measurable quantity. On the other hand, the maximum entropy (maxEnt) interpretation of statistical mechanics finds its basis in the information theoretic interpretation of entropy~\citep{shore,it_theory1,Press2012} where the inverse temperature $\beta$ is a Lagrange multiplier and entropy is an inference tool. Within the maxEnt framework $\beta$ can be treated as a parameter for the canonical ensemble distribution $P(r_{\bf B}|\beta)$. Fluctuations in $\beta$ can be interpreted as arising due to the hyper-probability $P(\beta)$ in the parameter space~\citep{caticha2004maximum}.

The maxEnt interpretation views statistical mechanics as an inference problem: amongst candidate distributions $P^*(r_{\bf B})$ that reproduce the known experimental measurements about the system ${\bf B}$, maxEnt states that the one that has the maximum entropy describes the internal states of the system correctly. For a system exchanging energy with the bath, the success of maxEnt is heavily dependent on the assumption of weak system-bath interactions (i.e. weak solute-solvent interactions)~\citep{it_theory1,Press2012}. Here, we have shown that the maxEnt interpretation also offers a way to generalize statistical mechanics. 

Note that regardless of the size of the system {\bf B}, the entropy of the joint system comprising of {\bf B} {\it and} the bath {\bf A} is maximized~\citep{lee:pre2012}. When the interactions between {\bf A} and {\bf B} are not too strong (but not too weak to be negligible either), we hypothesized that the effect of the variation in the states $r_{\bf A}$ of the solvent bath on the solute can be characterized by a varying its inverse temperature $\beta$. In other words, we assumed that the equivalence $P(r_{\bf A},r_{\bf B}) \leftrightarrow P(\beta,r_{\bf B})$ involved minimal loss of information about the bath. Our development lead to a generalization of statistical mechanics that resembles supestatistics~\citep{superstatistics,Crooks2008}. We established a direct correspondence between the developed framework and traditional statistical mechanics in the limiting case of very weak solute-solvent interactions. We also calculated the first order correction to canonical ensemble description of the state space of small systems.

We illustrated the  framework with a harmonic oscillator coupled to two solvent baths of very different chemical identities. We showed that the superstatistical distribution Eq.~\ref{eq:Pfinal1} describes the distribution of states $P(r)$ of the oscillator better than the usual canonical ensemble distribution Eq.\ref{eq:pr0}.  The current development has  shown that the thermodynamics of small systems can be suitably described by a superstatistics. We believe that the framework will be useful in understanding solvent induced modulations of solute state space e.g. with implicit solvent models.

\section{Acknowledgment}
I would like to thank Prof. Ken Dill, Prof. Dilip Asthagiri, Prof. German Drazer, and Mr. Sumedh Risbud  for stimulating conversations and suggestions about the manuscript.

This work was supported by grants PM-031 from the Office of Biological Research of the U.S. Department of Energy.

\bibliographystyle{biophysj}
\bibliography{superstat}

\pagebreak

\newpage

%
%
%
\section{Appendix}

\subsection{Numerical simulation}

A harmonic spring consisting of two Lennard-Jones particles was immersed in a bath of 512 Lennard-Jones particles in a cube of side 25\AA~and a bath of 256 TIP3~\citep{tip32,tip3mod} water molecules at 300. NVT molecular dynamics simulations were run with NAMD~\citep{namd:cc05} at 300K. The CHARMM~\citep{charmm:jpcb98} forcefield was used to describe the interaction between the Lennard-Jones particles and between the spring and the bath of particles.The spring constant for the harmonic oscillator was chosen to be $k=0.5$~kcal/mol$\cdot$\AA$^2$. The $\epsilon$ parameter for the Lennard-Jones bath bath was set at $-0.015$ while the $\epsilon$ parameter for the spring was set at $\epsilon=-10.0$ . The size parameter was set at $r = 2.1$\AA~for the oscillator particles ($r=1.1$\AA~ when interacting with water) and $r=1.1$\AA~for the bath particles. The systems were minimized for 2000 steps followed by an equilibration of 1 nanosecond and a production run of 2 nanosecond. Configurations were stored every 100 femtosecond.

\subsection{Fitting the parameters}

In order to fit Eq.~\ref{eq:Pfinal1}  to the experimental data, one needs to determine the free parameters from the simulation. In the traditional canonical ensemble, the inverse temperature $\beta$ of the harmonic oscillator will be estimated from its average energy. Here, we show how to estimate the free parameters from the simulation. It is non-trivial to measure the average system entropy $\langle S_{\bf B}(\bar \zeta) \rangle$ in a computer simulation. Yet, operationally,
\begin{eqnarray}
\langle S_{\bf B}(\beta) \rangle &\propto& \int \log \beta \cdot P(\beta) d\beta \nonumber \\
&\propto&\int \int \log r \cdot  P(r | \beta) P(\beta) dr d\beta \nonumber \\
&=& \int \langle \log r \rangle_{\beta} P(\beta) d\beta.
\end{eqnarray}
In other words, constraining $\overline {S_{\bf B}(\beta)}$ is equivalent to constraining $\log r$. Thus, in addition to $\langle E \rangle = \langle r^2 \rangle$, we also estimate  $\langle \log r \rangle$ from the simulation and then fit Eq.~\ref{eq:Pfinal1}.

\end{document}